\documentclass[pre,twocolumn,aps,10pt]{revtex4-2}
\usepackage{amsmath}
\usepackage{amssymb}
\usepackage{graphicx}
\usepackage{braket}
\usepackage{stmaryrd}
\usepackage{hyperref}
\usepackage{color}
\usepackage{dsfont}
\usepackage{ifthen}
\usepackage{bm}

\newboolean{showcomment}
\setboolean{showcomment}{true}

\hypersetup{
    colorlinks,
    citecolor=blue,
    linkcolor=blue,
    urlcolor=blue
}

\usepackage[whole]{bxcjkjatype}
\usepackage[normalem]{ulem}
\usepackage{color}
\usepackage{soul}
\definecolor{americanrose}{rgb}{1.0, 0.01, 0.24}
\definecolor{electricpurple}{rgb}{0.75, 0.0, 1.0}
\definecolor{vividgreen}{rgb}{0.0, 0.6, 0.0}
\definecolor{bittersweet}{rgb}{1.0, 0.44, 0.37}

\begin{document}


\title{Trade-off between Cooling-Step Count and Geometric Implementation Cost in Non-Markovian Algorithmic Cooling}

\author{Yohei Azumai}
\email{azumai@biom.t.u-tokyo.ac.jp}
\affiliation{Department of Information and Communication Engineering, Graduate
School of Information Science and Technology, The University of Tokyo,
Tokyo 113-8656, Japan}

\author{Yoshihiko Hasegawa}
\email{hasegawa@biom.t.u-tokyo.ac.jp}
\affiliation{Department of Information and Communication Engineering, Graduate
School of Information Science and Technology, The University of Tokyo,
Tokyo 113-8656, Japan}

\date{\today}
\begin{abstract} 

Quantum cooling is important for reliable quantum computation but involves a trade-off between cooling performance and implementation resources.
Although reservoir memory can improve particular aspects of cooling performance, the associated resource cost, particularly for circuit implementation, remains insufficiently understood.
Here, we investigate how reservoir memory affects the trade-off between cooling performance quantified by the cooling-step count and geometric implementation cost in single-qubit heat-bath algorithmic cooling.
Using a pseudomode mapping, we represent the non-Markovian damped Jaynes--Cummings dynamics by a repeated collision circuit and evaluate its geometric implementation cost.
Using matrix-based simulations and an implementation on the \texttt{ibm\_kawasaki} Heron r2 processor, we identify a trade-off: suppressing reservoir memory reduces the cooling-step count but generally increases the geometric protocol cost.
Our work provides a resource-based perspective on reservoir engineering for algorithmic cooling. 
\end{abstract}
\maketitle

\section{Introduction} 
\label{sec:Introduction} 
Reliable preparation of low-entropy states is essential for quantum computation. 
As quantum processors increase in size, thermal excitations and environmental noise degrade state initialization and subsequent quantum operations, making cooling and reset indispensable on current quantum hardware~\cite{Magnard2018,Aamir2025}. 
The third law of thermodynamics, however, precludes reaching absolute zero through a finite physical process, and even a small initialization temperature can substantially reduce the global fidelity of a large quantum register~\cite{Masanes2017,Scharlau2018,Buffoni2022}. 
Various cooling techniques, including laser cooling and dissipative many-body cooling, have therefore been developed to prepare quantum systems close to their ground states~\cite{Schreck2021,Scott2025,Li2025,Troyer2025, Lloyd2025}. 
Here, we focus on algorithmic cooling, in which quantum operations redistribute entropy among quantum subsystems to increase the polarization of a target qubit~\cite{Boykin2002}. 
In particular, heat-bath algorithmic cooling combines entropy compression with repeated thermal contact, allowing entropy to be removed from the computational system and enabling cooling beyond the closed-system limit~\cite{Rodriguez2017}. 

Quantum cooling generally involves a trade-off between cooling performance and the resources required to realize it. 
For example, improved cooling can require greater work, time, control complexity, or circuit complexity ~\cite{Oftelie2024,Taranto2023, Taranto:2025:CoolingGeometry}.
Early studies of heat-bath algorithmic cooling focused primarily on cooling performance, including achievable polarization bounds, improved cooling protocols, and the number of steps required to reach a prescribed polarization ~\cite{RodriguezBriones2016,Rodriguez2017,Alhambra2019, Clivaz2019}.
Moving beyond this performance-centered perspective, more recent studies have jointly evaluated cooling performance and resource-related quantities such as work cost, energy efficiency, finite-control complexity, gate complexity, and dissipation ~\cite{Clivaz2019E,Lin2024,Hu2025,Xuereb2025}.
These developments demonstrate the importance of accounting for performance-resource trade-offs when evaluating quantum cooling protocols, including heat-bath algorithmic cooling.
Existing studies of such performance-resource trade-offs have primarily considered closed-system unitary-control settings or open-system models without explicit reservoir memory.
Reservoir memory and non-Markovian effects, by contrast, have been shown to improve particular aspects of cooling performance ~\cite{Alhambra2019,Taranto2020,OrtegaTaberner2026}, but their resource implications have received considerably less attention.
In particular, although implementation-related resource costs have been investigated for cooling protocols based on Markovian collision models~\cite{Taranto:2025:CoolingGeometry}, an analogous circuit-level implementation cost for non-Markovian cooling dynamics remains insufficiently understood.
Consequently, it remains unclear how reservoir memory modifies the trade-off between cooling performance and implementation cost.

In this work, we address this question by investigating how reservoir memory affects the trade-off between cooling performance, quantified by the number of collision steps required to reach a prescribed cooling threshold, and the geometric implementation cost of single-qubit heat-bath algorithmic cooling. 
We describe the cooling dynamics using a damped Jaynes--Cummings model with a Lorentzian reservoir, which approximates optimal single-qubit heat-bath algorithmic cooling over a broad range of bath temperatures ~\cite{Alhambra2019}. 
The normalized spectral width \(\lambda/\gamma\) controls the reservoir-memory regime [cf. Eq.~\eqref{eq:spectral_density}]. 
Using the pseudomode representation~\cite{Garraway1997}, we map the non-Markovian reduced dynamics onto Markovian dynamics in an enlarged system and represent the resulting master equation by a repeated collision circuit [cf. Figs.~\ref{fig:pseudomode} and \ref{fig:collision_step}]. 
To quantify its implementation cost, we adopt a dilation-based geometric formulation for open quantum channels~\cite{Nielsen2006,Chapman2021,Tan2026,Acevedo2025, Acevedo:2026:GeometricChannels} and evaluate the path length accumulated until the target reaches a prescribed cooling threshold [cf. Eq.~\eqref{eq:Cgeom_cool} and Fig.~\ref{fig:geometric_channel_cost}].
We evaluate the cooling-step count and geometric protocol cost using matrix-based simulations and an implementation on the \texttt{ibm\_kawasaki} Heron r2 processor. 
Over the parameter range considered here, the cooling-step count decreases as reservoir memory is suppressed, whereas the geometric protocol cost generally increases because the geometric length of each collision step becomes larger [cf. Figs.~\ref{fig:matrix_Ncool}--\ref{fig:hardware_Cgeom}]. 
The hardware measurements qualitatively reproduce these contrasting dependences despite device noise and incomplete threshold crossings. 
These results identify a trade-off in the specified circuit implementation: suppressing reservoir memory reduces the number of steps required for cooling but increases the geometric implementation cost accumulated until the cooling threshold is reached.

\section{Methods}
\label{sec:Methods}

\subsection{Damped Jaynes--Cummings model with a Lorentzian reservoir}
\label{subsec:JC_model}

We consider a two-level target system \(S\) coupled to a structured bosonic reservoir. 
This choice is motivated by its connection to single-qubit algorithmic cooling. 
Alhambra et al.~\cite{Alhambra2019} showed that, for a single target qubit, an optimal heat-bath algorithmic-cooling protocol can be well approximated over a broad range of bath temperatures by a Jaynes--Cummings interaction between the target qubit and a single thermal bosonic mode. 
The model also provides a physically accessible setting in which reservoir memory can be exploited to improve cooling. 
We therefore adopt the damped Jaynes--Cummings model as a microscopic cooling primitive and investigate the circuit resources required to implement its dynamics.

The total system-reservoir Hamiltonian is
\begin{align}
    H_{\rm JC}
    =
    \omega_0 \sigma_+\sigma_-
    +
    \sum_k \omega_k b_k^\dagger b_k
    +
    \sum_k
    \left(
    g_k\sigma_+b_k
    +
    g_k^*\sigma_-b_k^\dagger
    \right),
    \label{eq:H_JC}
\end{align}
where \(\omega_0\) is the transition frequency of the target system, 
\(\sigma_+\) and \(\sigma_-\) are its raising and lowering operators, 
\(b_k^\dagger\) and \(b_k\) are the creation and annihilation operators, respectively, of the \(k\)-th reservoir mode with frequency \(\omega_k\), and \(g_k\) is the corresponding system-reservoir coupling strength.

We consider a reservoir with the Lorentzian spectral density:
\begin{align}
    J(\omega)
    =
    \frac{1}{2\pi}
    \frac{\gamma\lambda^2}
    {(\omega_0-\omega-\delta)^2+\lambda^2},
    \label{eq:spectral_density}
\end{align}
where \(\gamma>0\) characterizes the overall system-reservoir coupling scale, 
\(\lambda>0\) is the spectral width, and
\(\delta=\omega_0-\omega_c\) is the detuning between the system transition frequency \(\omega_0\) and the central reservoir frequency \(\omega_c\).

The Fourier transform of Eq.~\eqref{eq:spectral_density} gives the reservoir correlation function
\(f(t-t')=(\gamma\lambda/2)\exp[-(\lambda-i\delta)(t-t')]\) for \(t\geq t'\). 
Its envelope decays on the characteristic timescale \(\tau_E\sim\lambda^{-1}\). 
A narrow Lorentzian spectrum therefore produces long-lived reservoir correlations, whereas a broad spectrum rapidly loses its memory and approaches the Markovian limit.
In this work, we use \(\lambda\) as a controllable reservoir-memory parameter: decreasing \(\lambda\) increases the duration of the reservoir correlations, whereas increasing \(\lambda\) approaches the Markovian limit. 
This interpretation is consistent with previous analyses of the damped Jaynes--Cummings model, in which the Lorentzian spectral width controls information backflow and the transition between Markovian and non-Markovian dynamical regimes~\cite{Shahri2023}.

For the parameter range studied here, we further support this reservoir-memory classification using the Breuer--Laine--Piilo information-backflow measure~\cite{Breuer2009,Laine2010}. As detailed in Appendix~\ref{app:BLP}, the BLP measure is evaluated from the continuous damped Jaynes--Cummings dynamics over a fixed long-time interval. 
Its dependence on \(\lambda/\gamma\) provides a complementary characterization of the more non-Markovian and more Markovian sides of the reservoir family considered here.

\subsection{Pseudomode representation}
\label{subsec:pseudomode}

The pseudomode method maps the non-Markovian dynamics generated by a structured reservoir onto Markovian dynamics in an enlarged Hilbert space~\cite{Garraway1997}. 
For a Lorentzian spectral density, the reservoir memory can be represented by a single damped pseudomode \(P\), while the remaining background reservoir is Markovian. 
This construction has been widely used to analyze structured open-system dynamics and has recently been extended and applied in a variety of non-Markovian settings~\cite{GallinaPseudomode,ZhangPseudomode}.

\begin{figure*}
    \centering
    \includegraphics[width=0.9\textwidth]{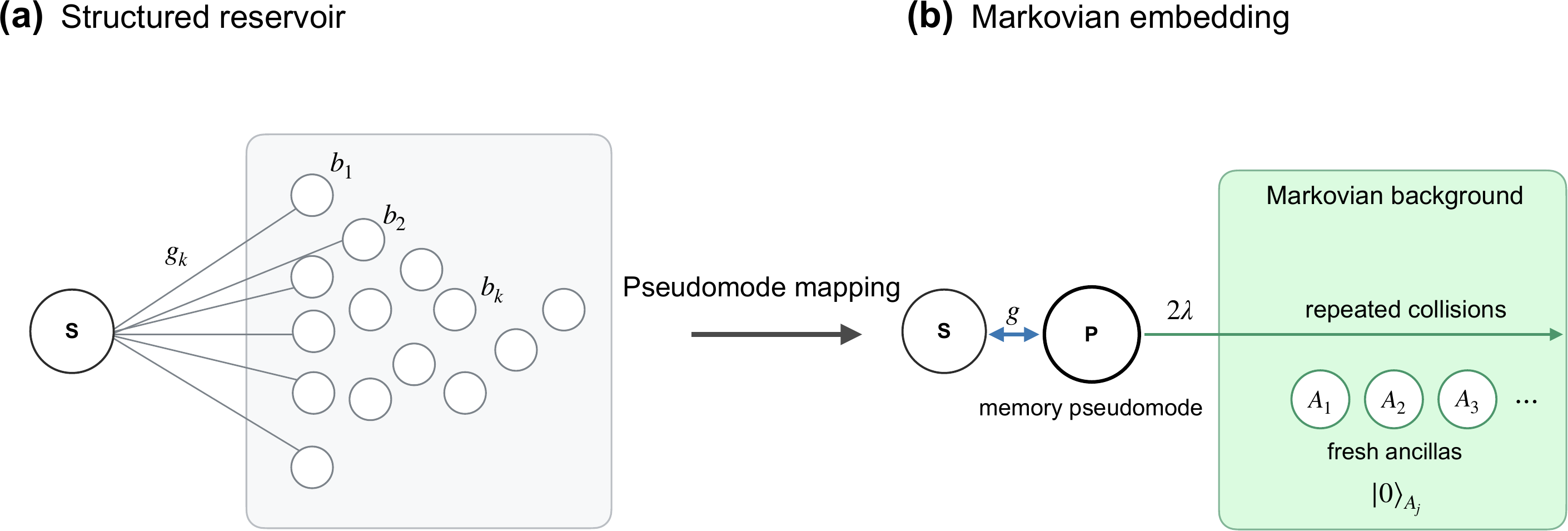}
    \caption{
    Pseudomode representation of the Lorentzian reservoir. (a) The target system \(S\) interacts with a structured bosonic reservoir composed of modes \(b_k\). (b) Under the pseudomode mapping, the reservoir memory is represented by a pseudomode \(P\), which coherently exchanges excitations with \(S\), while the residual Markovian background is implemented by repeated interactions with fresh ancillas \(A_j\).
    }
    \label{fig:pseudomode}
\end{figure*}

In the present model, we regard \(S+P\) as the enlarged system. 
The target qubit \(S\) coherently exchanges excitations with \(P\), which stores the reservoir memory, while \(P\) decays irreversibly into a Markovian background. 
In the collision-circuit implementation introduced below, this background is represented by a sequence of fresh ancillas \(A\), as illustrated in Fig.~\ref{fig:pseudomode}.

In the single-excitation sector, the pseudomode can be represented as an effective two-level degree of freedom. 
The dynamics of the enlarged density matrix \(\rho_{SP}\) is then governed by the Markovian master equation:
\begin{align}
    \frac{d\rho_{SP}}{dt}
    =
    -i
    \left[
    H_{SP}+H_\delta,
    \rho_{SP}
    \right]
    +
    2\lambda
    \mathcal{D}[\sigma_-^P]\rho_{SP},
    \label{eq:pseudomode_master}
\end{align}
where
\(H_{SP}
=
g(\sigma_+^S\sigma_-^P+\sigma_-^S\sigma_+^P)\)
describes the coherent excitation exchange,
\(g=\sqrt{\gamma\lambda/2}\) is the pseudomode coupling strength,
and \(H_\delta=-\delta n_P\), with
\(n_P=\sigma_+^P\sigma_-^P\), accounts for the detuning.
The dissipator is defined as
\(\mathcal{D}[L]\rho
=
L\rho L^\dagger
-\{L^\dagger L,\rho\}/2\).
Thus, the original non-Markovian dynamics of \(S\) is recovered by tracing out the pseudomode from the Markovian dynamics of \(S+P\).

\subsection{Discrete collision circuit}
\label{subsec:collision_circuit}

The pseudomode representation recasts the original non-Markovian reduced dynamics as the Markovian master equation for the enlarged system \(S+P\) in Eq.~\eqref{eq:pseudomode_master}. 
Within the collision-model framework, the residual Markovian decay can be represented by repeated interactions with fresh ancillas, providing a circuit implementation of the cooling dynamics and enabling us to evaluate its geometric implementation cost.
Accordingly, we represent Eq.~\eqref{eq:pseudomode_master} by a repeated collision circuit.
The construction follows the general pseudomode-circuit strategy of
Ref.~\cite{GallinaPseudomode}, in which the environmental memory is
represented explicitly by a pseudomode and its residual Markovian
decay is reproduced through collisions with fresh ancillas.
Whereas Ref.~\cite{GallinaPseudomode} considered dephasing dynamics,
the present construction is adapted to the dissipative
Jaynes--Cummings model relevant to cooling, involving excitation
exchange between the target system, pseudomode, and ancillas.

\begin{figure*}
    \centering
    \includegraphics[width=0.9\textwidth]
    {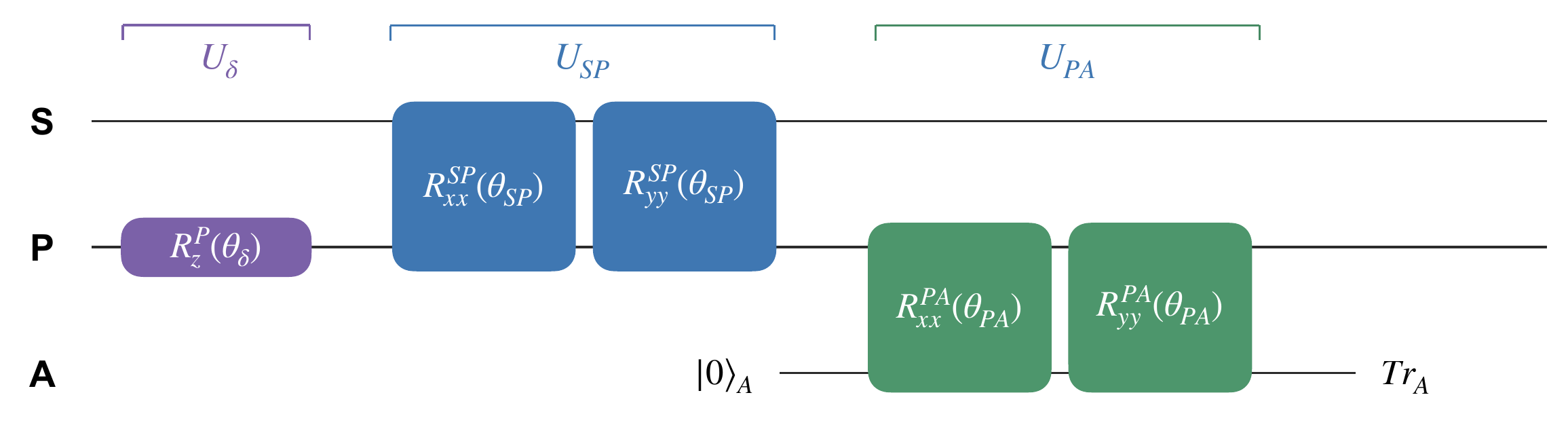}
    \caption{
    Gate decomposition of one pseudomode collision step. The detuning, system-pseudomode exchange, and pseudomode-ancilla collision operations are grouped as \(U_\delta\), \(U_{SP}\), and \(U_{PA}\), respectively. The ancilla is initialized in \(\lvert0\rangle_A\), traced out after the collision, and replaced by a fresh ancilla at the next step. The explicit construction, continuous-time limit, and numerical validation are given in Appendix~\ref{app:continuous_limit}.
    }
    \label{fig:collision_step}
\end{figure*}

Figure~\ref{fig:collision_step} shows one collision step.
The pseudomode first undergoes a detuning rotation and then exchanges
an excitation coherently with the target system.
It subsequently interacts with an ancilla initialized in
\(\lvert0\rangle_A\).
After the collision, the ancilla is traced out and replaced by a fresh
ancilla at the next step.

Repeated application of this step defines the discrete cooling
dynamics.
The rotation angles are chosen such that the circuit converges to the
pseudomode master equation in Eq.~\eqref{eq:pseudomode_master} as
\(\Delta t\rightarrow0\).
The explicit one-step channel, its continuous-time limit, and the numerical validation against the continuous pseudomode dynamics are given in Appendix~\ref{app:continuous_limit}.

\subsection{Geometric protocol cost}
\label{subsec:complexity}

We quantify the implementation cost of the cooling protocol by the
geometric length accumulated until the target system reaches a
prescribed cooling threshold. Specifically, we define
\begin{align}
    \mathcal{C}_{\rm geom}^{\rm cool}(\lambda;\epsilon)
    =
    N_{\rm cool}(\lambda;\epsilon)
    \ell_{\rm PM}^{\rm seq}(\lambda),
    \label{eq:Cgeom_cool}
\end{align}
where \(N_{\rm cool}(\lambda;\epsilon)\) is the number of collision
steps required to reach the threshold \(\epsilon\) at spectral width
\(\lambda\), and \(\ell_{\rm PM}^{\rm seq}(\lambda)\) is the
corresponding geometric length of one collision step.

The geometric construction follows the dilation-based channel
complexity introduced in
Refs.~\cite{Acevedo2025,Acevedo:2026:GeometricChannels},
which extends the geometric approach to closed-system circuit
complexity developed by Nielsen and collaborators~\cite{Nielsen2006}.
Nielsen's approach reformulates the problem of synthesizing a target unitary as a shortest-path problem on the Riemannian manifold of unitary operators.
With an appropriately chosen metric, the geodesic distance from the identity to the target unitary serves as a continuous measure closely related to the minimum gate count required to implement it.

\begin{figure*}
    \centering 
    \includegraphics[width=0.9\textwidth] 
    {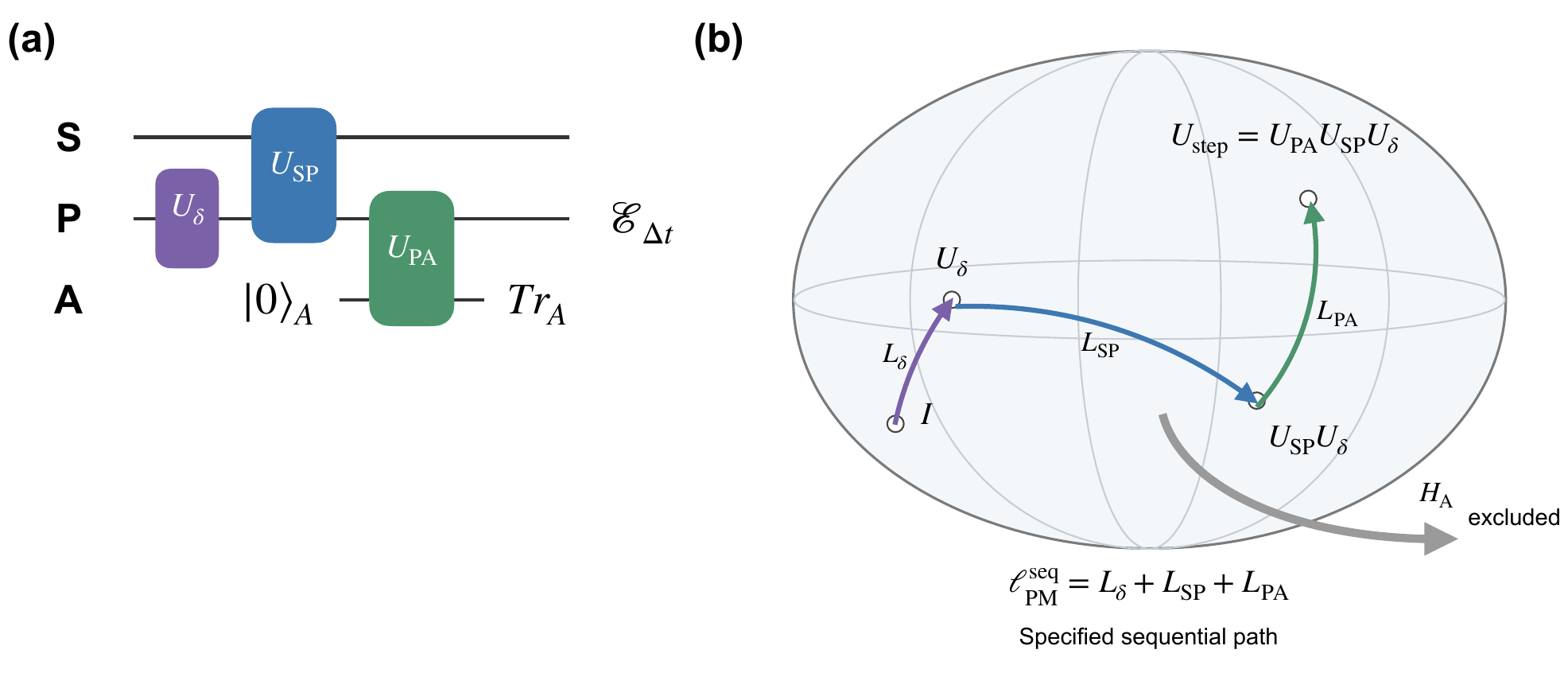} 
    \caption{ 
    Geometric representation of one pseudomode collision step. (a) The sequential circuit implements \(U_{\rm step}=U_{PA}U_{SP}U_\delta\) on the enlarged system \(\widetilde{S}=S+P\) and a fresh ancilla \(A\), which is traced out after the collision. 
    (b) The detuning, system-pseudomode exchange, and pseudomode-ancilla collision define consecutive segments of the specified unitary path, with lengths \(L_\delta\), \(L_{SP}\), and \(L_{PA}\), respectively. Their sum gives \(\ell_{\rm PM}^{\rm seq}=L_\delta+L_{SP}+L_{PA}\). The depicted path is not minimized over alternative circuit paths or unitary dilations. 
    Details are given in Appendix~\ref{app:geometric_cost}.
    } 
    \label{fig:geometric_channel_cost} 
\end{figure*}

As illustrated in Fig.~\ref{fig:geometric_channel_cost}, one collision step is represented as a piecewise unitary path on \(\widetilde{S}+A\), with \(\widetilde{S}=S+P\), and its one-step length is obtained by summing the geometric lengths of the sequential circuit segments. Generator components that act only on the ancilla are excluded because they do not affect the reduced dynamics of the enlarged system.
For the present analysis, we adopt the Hilbert--Schmidt specialization~\cite{Acevedo:2026:GeometricChannels}, in which equal penalty weights are assigned to all generator directions, as an analytically convenient baseline.
With this choice, the unitary group equipped with the corresponding metric forms a homogeneous Riemannian manifold, and all normalized Pauli-string generator directions are assigned the same implementation cost.
The framework also permits anisotropic penalties, which may be used to incorporate hardware-dependent control costs.

The cooling-step count is defined by
\begin{align}
    N_{\rm cool}(\lambda;\epsilon)
    =
    \min
    \left\{
    N:
    \operatorname{Tr}
    \left[
    n_S\rho_S(N\Delta t;\lambda)
    \right]
    \leq
    \epsilon
    \right\},
    \label{eq:Ncool_and_Pe}
\end{align}
where \(n_S=\sigma_+^S\sigma_-^S\) is the excitation-number operator of the target system.

For the sequential collision circuit shown in
Fig.~\ref{fig:collision_step}, the resulting one-step geometric length
is
\begin{align}
    \ell_{\rm PM}^{\rm seq}(\lambda)
    =
    \sqrt{\frac{8}{63}}
    \left[
    \sqrt{\frac{\gamma\lambda}{2}}\Delta t
    +
    \sqrt{2\lambda\Delta t}
    \right]
    +
    \sqrt{\frac{2}{63}}
    |\delta|\Delta t.
    \label{eq:ellPM_lambda}
\end{align}
The derivation of Eq.~\eqref{eq:ellPM_lambda}, including the
environment-only projection and the contribution of each circuit
segment, is given in Appendix~\ref{app:geometric_cost}.

The quantity in Eq.~\eqref{eq:Cgeom_cool} is the
implementation-dependent geometric length of the specified sequential
collision circuit. It is not minimized over all possible circuit
paths or all possible dilations of the reduced cooling channel and
therefore should not be identified with the intrinsic channel
complexity.


\section{Numerical Results}
\label{sec:Results}

To examine the relation between non-Markovianity and the geometric
protocol cost defined in Eq.~\eqref{eq:Cgeom_cool}, we vary the
dimensionless spectral width \(\lambda/\gamma\) and evaluate
\(N_{\rm cool}(\lambda;\epsilon)\) and
\(\mathcal{C}_{\rm geom}^{\rm cool}(\lambda;\epsilon)\).
As discussed in Sec.~\ref{subsec:JC_model},
\(\lambda/\gamma\) controls the reservoir-memory regime: a smaller
value corresponds to a more non-Markovian reservoir, whereas
a larger value corresponds to a more Markovian reservoir.
The long-time BLP calculation in Appendix~\ref{app:BLP} further
supports the reservoir-memory classification over the spectral-width
range considered here.
We first perform matrix-based density-matrix propagation and then
implement the same spectral-width sweep on the \texttt{ibm\_kawasaki} Heron r2 processor.

\subsection{Matrix-based simulation}
\label{subsec:matrix_results}

\begin{figure}
    \centering
    \includegraphics[width=0.9\linewidth]
    {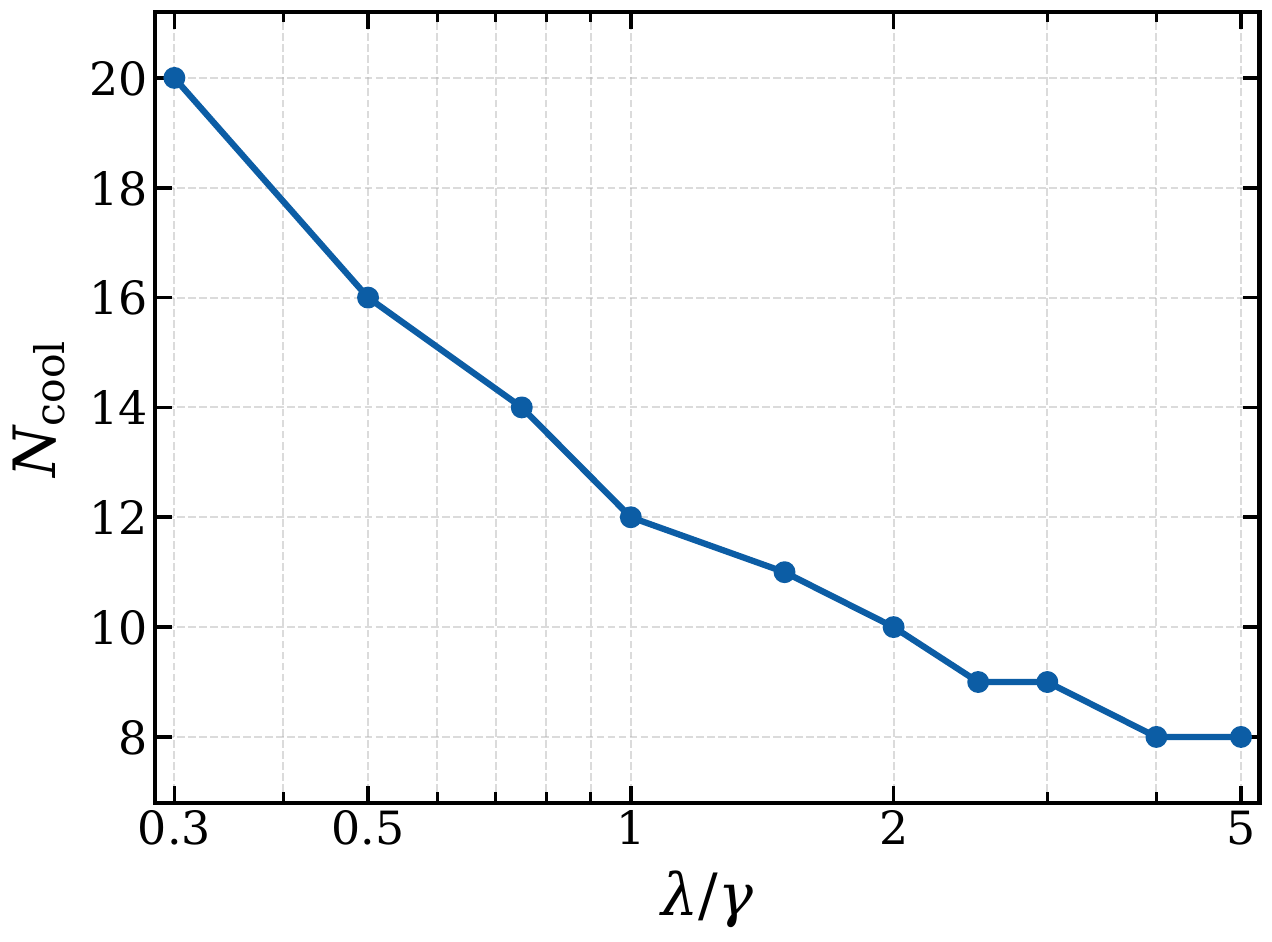}
    \caption{
    Cooling-step count \(N_{\rm cool}(\lambda;\epsilon)\) obtained from matrix-based density-matrix propagation as a function of the normalized spectral width \(\lambda/\gamma\). The analysis is restricted to \(0.3\leq\lambda/\gamma\leq5\), over which the collision circuit reliably approximates the continuous pseudomode dynamics, as verified in Appendix~\ref{app:continuous_limit}. The target is initialized with the Bloch vector \((0.5,0.5,0.5)\). The parameters are \(\gamma\Delta t=0.1\), \(\delta=0\), \(\epsilon=0.15\), and \(N_{\max}=20\).
    The line is a guide for the eye.
    }
    \label{fig:matrix_Ncool}
\end{figure}

Figure~\ref{fig:matrix_Ncool} shows
\(N_{\rm cool}(\lambda;\epsilon)\) as a function of
\(\lambda/\gamma\).
On the narrow-band, more non-Markovian side, \(N_{\rm cool}\) is large. 
Over the range considered here, the cooling-step count decreases as the spectral width is increased. 
Thus, suppressing reservoir memory reduces the number of collision steps required to reach the cooling threshold.

\begin{figure}
    \centering
    \includegraphics[width=0.9\linewidth]
    {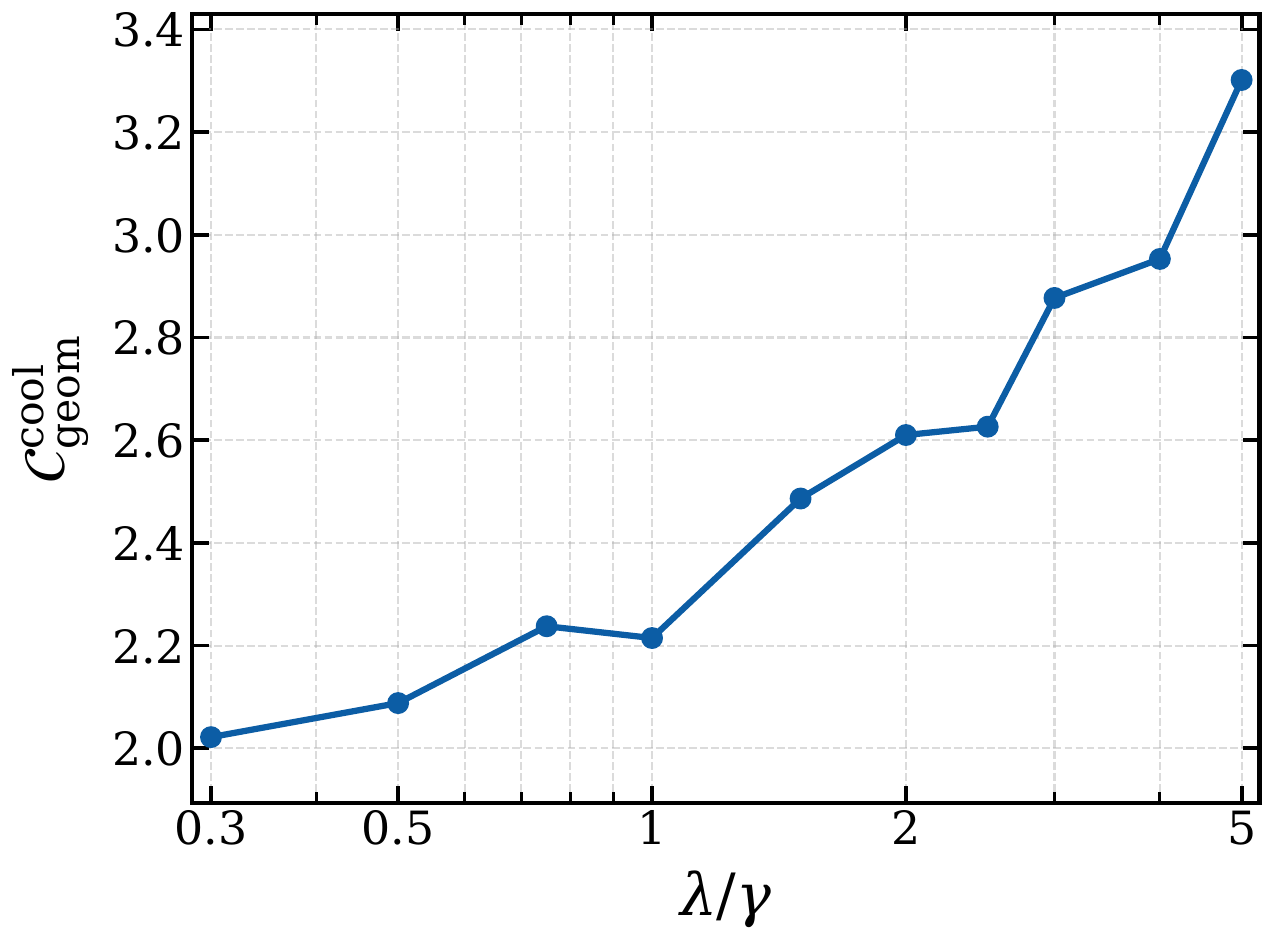}
    \caption{
    Accumulated geometric cooling cost
    \(\mathcal{C}_{\rm geom}^{\rm cool}(\lambda;\epsilon)\) obtained
    from the matrix-based simulation.
    The cost is evaluated from
    Eq.~\eqref{eq:Cgeom_cool} using
    \(N_{\rm cool}(\lambda;\epsilon)\) and the one-step geometric
    length \(\ell_{\rm PM}^{\rm seq}(\lambda)\).
    The parameters are the same as in
    Fig.~\ref{fig:matrix_Ncool}. The line is a guide for the eye.
    }
    \label{fig:matrix_Cgeom}
\end{figure}

Figure~\ref{fig:matrix_Cgeom} shows the corresponding accumulated
geometric cost
\(\mathcal{C}_{\rm geom}^{\rm cool}(\lambda;\epsilon)\).
In contrast to \(N_{\rm cool}\), the geometric cost generally increases with \(\lambda/\gamma\) over the range considered here, with small stepwise variations arising from the discrete definition of \(N_{\rm cool}\).
This overall behavior originates from the \(\lambda\) dependence of the
rotation angles in Eq.~\eqref{eq:ellPM_lambda}: decreasing
\(\lambda\) reduces the geometric length
\(\ell_{\rm PM}^{\rm seq}(\lambda)\) of each collision step.
Overall, the decrease in the one-step length outweighs the larger number of collisions required on the narrow-band side.
Therefore, suppressing reservoir memory generally reduces the cooling-step count but increases the geometric implementation cost of the specified collision circuit.

\subsection{Hardware implementation}
\label{subsec:hardware_results}

\begin{figure}
    \centering
    \includegraphics[width=0.9\linewidth]
    {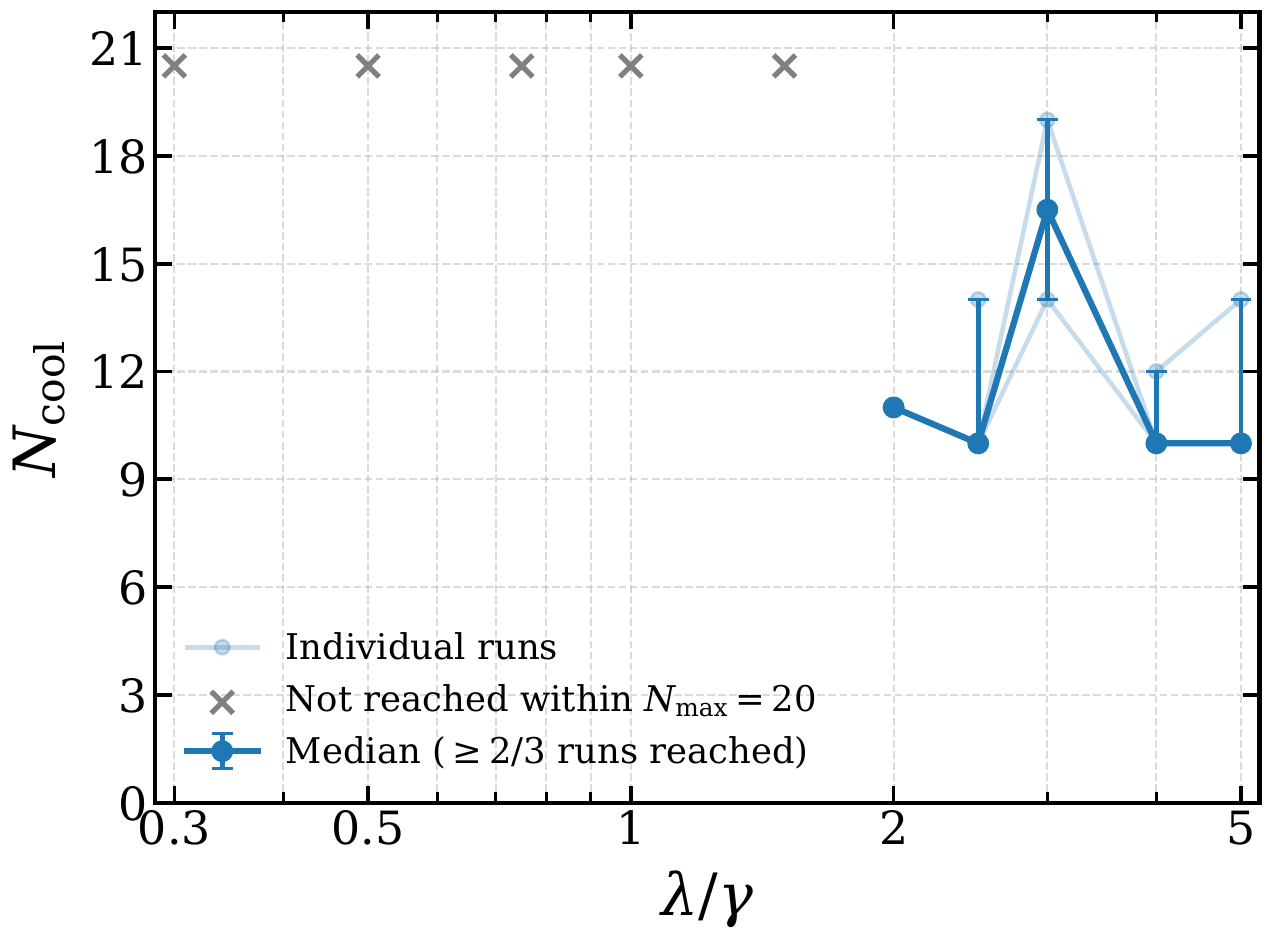}
    \caption{
    Cooling-step count \(N_{\rm cool}(\lambda;\epsilon)\) measured on
    the \texttt{ibm\_kawasaki} Heron r2 processor.
    Thin curves show three independent hardware runs.
    The thick curve shows the median at points where the threshold was
    reached in at least two of the three runs, and the error bars span the corresponding minimum and maximum values.
    Crosses denote points where the threshold was not reached in any
    run within \(N_{\max}=20\).
    The target is initialized with the Bloch vector
    \((0.5,0.5,0.5)\).
    The analysis is restricted to \(0.3\leq\lambda/\gamma\leq5\), and the parameters are
    \(\gamma\Delta t=0.1\),
    \(\delta=0\),
    \(\epsilon=0.15\), and \(8192\) shots per circuit.
    Dynamical decoupling is applied to the repeated-reset circuit. The lines are guides for the eye.
    }
    \label{fig:hardware_Ncool}
\end{figure}

We next implement the collision circuit on the
\texttt{ibm\_kawasaki} Heron r2 processor.
A single collision ancilla is reset between successive steps, and
dynamical decoupling is applied to suppress decoherence during the
reset-induced idle intervals.
Figure~\ref{fig:hardware_Ncool} shows the results of three independent
hardware runs.
The plotted representative values are the medians of the runs that
reach the cooling threshold, provided that the threshold is reached
in at least two of the three runs.
On the smaller-\(\lambda/\gamma\), more non-Markovian side, the threshold is frequently not reached within \(N_{\max}=20\), indicating that a larger number of collision steps may be required. 
By contrast, on the larger-\(\lambda/\gamma\), more Markovian side, the threshold is reached within the observation window.
Although hardware noise and run-to-run fluctuations prevent a monotonic decrease in \(N_{\rm cool}\) from being resolved, the contrast between the smaller- and larger-\(\lambda/\gamma\) sides is consistent with the matrix-based simulation.

\begin{figure}
    \centering
    \includegraphics[width=0.9\linewidth]
    {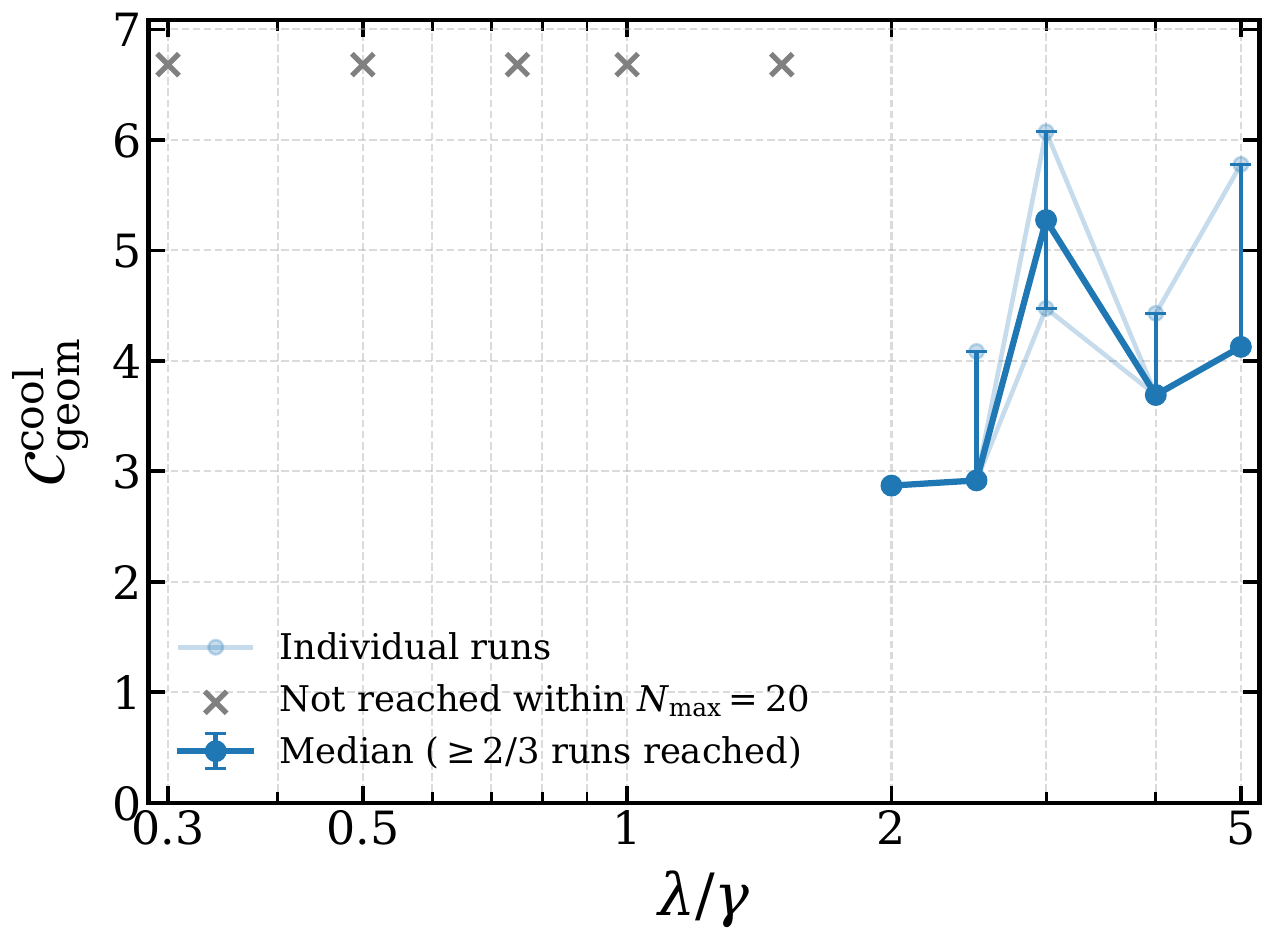}
    \caption{
    Geometric cooling cost
    \(\mathcal{C}_{\rm geom}^{\rm cool}(\lambda;\epsilon)\)
    calculated from the hardware cooling-step measurements.
    Thin curves show the individual runs.
    The thick curve shows the median at points where the cooling
    threshold was reached in at least two of the three runs, and the
    error bars span the corresponding minimum and maximum values.
    Crosses denote points where the threshold was not reached in any run within \(N_{\max}=20\).
    The cost is calculated using the definition in
    Eq.~\eqref{eq:Cgeom_cool}.
    The experimental parameters are the same as in
    Fig.~\ref{fig:hardware_Ncool}. The lines are guides for the eye.
    }
    \label{fig:hardware_Cgeom}
\end{figure}

Figure~\ref{fig:hardware_Cgeom} shows the hardware geometric cost
calculated from the measured \(N_{\rm cool}(\lambda;\epsilon)\) using
the definition in Eq.~\eqref{eq:Cgeom_cool}.
The cost is defined only at spectral widths for which the cooling threshold is reached within \(N_{\max}=20\).
Among these spectral widths, the geometric cost and its run-to-run variation are smaller on the smaller-\(\lambda/\gamma\) side.
Although hardware noise prevents a monotonic dependence from being resolved, this behavior is qualitatively consistent with the matrix-based simulation, in which stronger reservoir memory reduces the geometric implementation cost of the specified collision circuit.
Taken together with the hardware cooling-step results, these observations qualitatively reproduce the trade-off found in the matrix-based simulation: weaker reservoir memory reduces the number of cooling steps but increases the geometric implementation cost.
Finally, to corroborate the cooling-performance trend inferred from \(N_{\rm cool}\), we evaluate the minimum excited-state population attained within the fixed observation window. 
As shown in Appendix~\ref{app:min_population}, this complementary measure yields the same broad conclusion for both the matrix-based simulation and hardware implementation: under the present protocol and parameter range, better finite-step cooling performance is generally obtained on the larger-\(\lambda/\gamma\), more Markovian side.

\section{Conclusion}
\label{sec:Conclusion}

In this work, we investigated how reservoir non-Markovianity affects the trade-off between the cooling-step count and the geometric implementation cost of single-qubit heat-bath algorithmic cooling.
We described the cooling dynamics using a damped Jaynes--Cummings model with a Lorentzian reservoir, mapped the structured environment to a damped pseudomode, and represented the resulting dynamics by a repeated collision circuit.
For this circuit, we derived an implementation-dependent geometric
protocol cost from the unitary dilation path and examined its
dependence on the normalized spectral width \(\lambda/\gamma\).
Over the parameter range considered here, the cooling-step count decreased as the reservoir memory was suppressed, whereas the geometric protocol cost generally increased because the geometric length of each collision step became larger.
We implemented the collision circuit on the
\texttt{ibm\_kawasaki} Heron r2 processor and observed qualitative
spectral-width dependences consistent with the discrete-circuit
simulation, despite hardware noise and incomplete threshold crossings
at some parameter values.
Overall, these results identify a trade-off in the specified circuit implementation: suppressing reservoir memory reduces the number of steps required for cooling but increases the geometric implementation cost accumulated until the cooling threshold is reached.

\begin{acknowledgments}

This work was supported by JSPS KAKENHI Grant Number JP23K24915. 

We acknowledge the use of IBM Quantum services for this work. The views expressed are those of the authors, and do not reflect the official policy or position of IBM or the IBM Quantum team.

Microsoft 365 Copilot, based on the GPT-5 reasoning model, was used to assist with the generation, revision, and debugging of numerical simulation code and with selected checks of calculations and derivations. The research questions, physical models, parameter choices, and interpretation of the results were determined by the authors. All AI-assisted code and outputs were reviewed by the authors and verified through direct code inspection, analytical consistency checks, and numerical validation where applicable. The authors take full responsibility for the calculations, numerical results, and scientific conclusions presented in this work.

\end{acknowledgments}

\appendix

\section{BLP non-Markovianity of the damped Jaynes--Cummings model} 
\label{app:BLP}

The reservoir correlation time \(\tau_E\sim\lambda^{-1}\) provides a physical estimate of the duration of the environmental memory. 
Several measures have been proposed to characterize non-Markovian quantum dynamics~\cite{Addis2014}. 
Here, to complement the correlation-time characterization, we use the Breuer--Laine--Piilo (BLP) measure based on information backflow~\cite{Breuer2009,Laine2010}, which has also been applied to quantify non-Markovianity in quantum refrigeration~\cite{Mondkar2026}. 
For two initial system states \(\rho_1(0)\) and \(\rho_2(0)\), their distinguishability at time \(t\) is quantified by the trace distance:
\begin{align} 
D\left( \rho_1(t),\rho_2(t) \right) = \frac{1}{2} \left\| \rho_1(t)-\rho_2(t) \right\|_1. \label{eq:BLP_trace_distance} 
\end{align}
The corresponding information-flow rate is 
\begin{align}
\sigma \left( t,\rho_1(0),\rho_2(0) \right) = \frac{d}{dt} D\left( \rho_1(t),\rho_2(t) \right). 
\label{eq:BLP_information_rate}
\end{align} 
A temporary increase in trace distance, \(\sigma(t,\rho_1(0),\rho_2(0))>0\), indicates information backflow from the environment to the system. 
The BLP measure is therefore defined as 
\begin{align} 
\mathcal{N}_{\rm BLP} = \max_{\rho_1(0),\rho_2(0)} \int_{\sigma>0} dt\, \sigma \left( t,\rho_1(0),\rho_2(0) \right). 
\label{eq:BLP_measure} 
\end{align} 

For the amplitude-damping dynamics of the damped Jaynes--Cummings model, previous numerical analyses based on extensive sampling of initial-state pairs indicate that the maximum in Eq.~\eqref{eq:BLP_measure} is attained for the ground- and excited-state pair, \(\rho_1(0)=\lvert0\rangle\langle0\rvert\) and \(\rho_2(0)=\lvert1\rangle\langle1\rvert\)~\cite{Breuer2009,Shahri2023}.
For the Lorentzian spectral density considered here, the target-system amplitude is
\begin{align} 
    c_S(t) = e^{-(\lambda-i\delta)t/2} \left[ \cosh\left(\frac{\Omega t}{2}\right) + \frac{\lambda-i\delta}{\Omega} \sinh\left(\frac{\Omega t}{2}\right) \right], 
\label{eq:BLP_cS} 
\end{align} 
where \(\Omega=\sqrt{(\lambda-i\delta)^2-2\gamma\lambda}\). 
For this maximizing pair, the trace distance defined in Eq.~\eqref{eq:BLP_trace_distance} becomes
\begin{align} 
    D_{\rm opt}(t) = \left|c_S(t)\right|^2.
    \label{eq:BLP_optimal_distance} 
\end{align}
Using Eq.~\eqref{eq:BLP_optimal_distance}, we approximate the positive-rate integral in Eq.~\eqref{eq:BLP_measure} as
\begin{align} 
    \mathcal{N}_{\rm BLP}^{(\Delta t_{\rm BLP})} 
    &= 
    \sum_n \max \left[ 0, \frac{ D_{\rm opt}(t_{n+1})-D_{\rm opt}(t_n) }{ \Delta t_{\rm BLP} } \right] \Delta t_{\rm BLP} 
    \nonumber\\
    &= 
    \sum_n \max \left[ 0, \left|c_S(t_{n+1})\right|^2 - \left|c_S(t_n)\right|^2 \right],
    \label{eq:BLP_discrete}
\end{align} 
where \(t_n=n\Delta t_{\rm BLP}\). 
To capture the long-time information-backflow behavior, the measure is evaluated over the fixed dimensionless interval \(0\leq\gamma t\leq100\) with \(\gamma\Delta t_{\rm BLP}=10^{-3}\).
The calculation is performed on a dense spectral-width grid over \(0.3\leq\lambda/\gamma\leq5\), under the same zero-detuning condition as in the main analysis.

\begin{figure} 
    \centering 
    \includegraphics[width=0.9\linewidth] {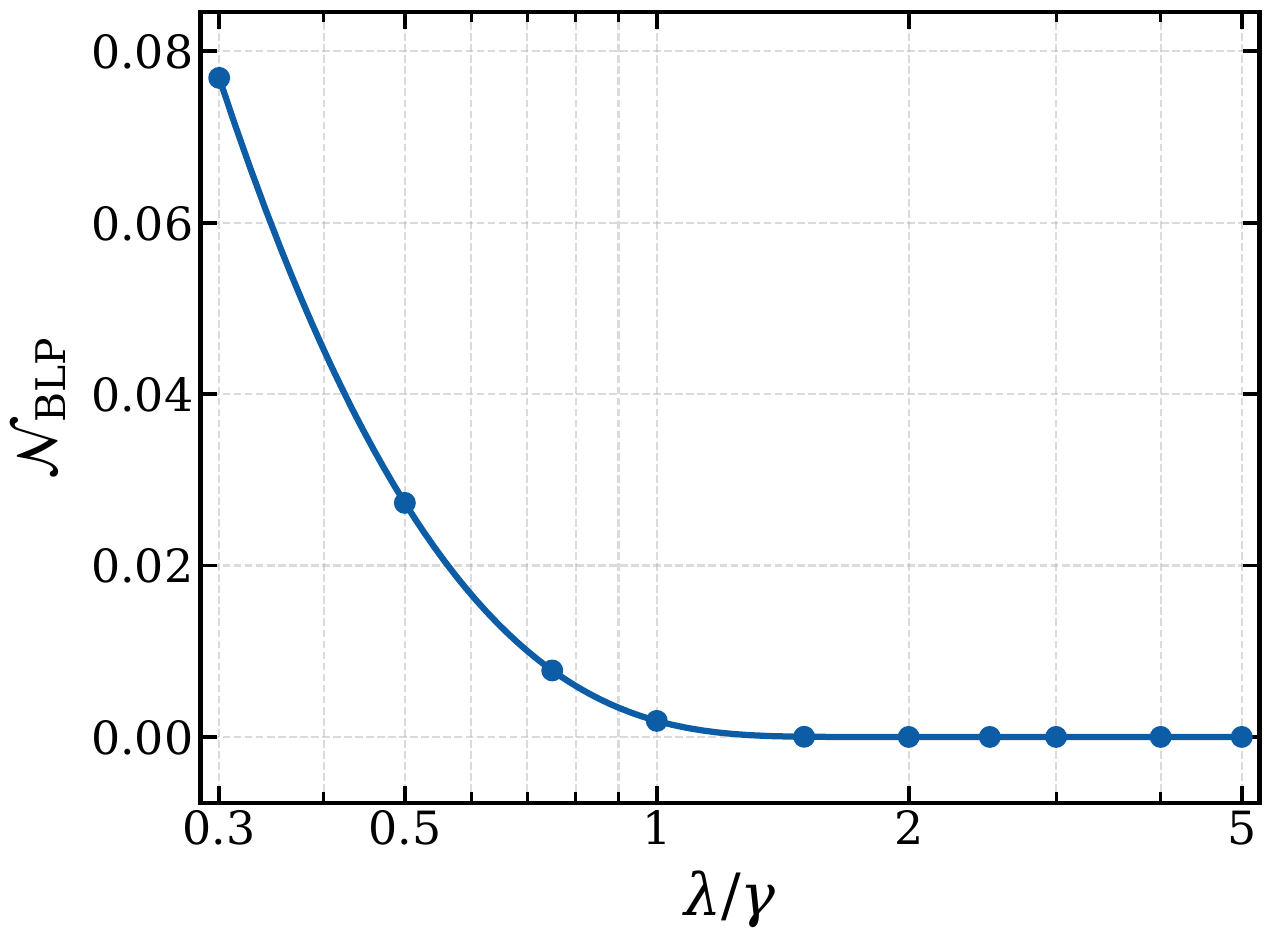} 
    \caption{
    BLP non-Markovianity measure for the continuous damped Jaynes--Cummings dynamics as a function of the normalized Lorentzian spectral width \(\lambda/\gamma\).
    The measure is evaluated from the trace-distance revival of the excited- and ground-state pair over the fixed dimensionless interval \(0\leq\gamma t\leq100\), with \(\delta=0\). 
    The solid curve shows the numerical evaluation over a dense spectral-width grid in \(0.3\leq\lambda/\gamma\leq5\), while the markers indicate the spectral widths used in the matrix-based simulation and hardware implementation in the main text. 
    A positive value indicates information backflow.
    The measure becomes nearly zero on the larger-\(\lambda/\gamma\) side, supporting the classification of this side as more Markovian, whereas the positive values on the narrow-band side support its classification as more non-Markovian.} 
\label{fig:BLP_lambda} 
\end{figure}

Figure~\ref{fig:BLP_lambda} shows the resulting BLP measure as a function of \(\lambda/\gamma\). 
The BLP measure becomes nearly zero on the larger-\(\lambda/\gamma\) side of the considered range, indicating the suppression of information backflow. By contrast, its positive values on the narrow-band side indicate non-Markovianity.
This behavior is consistent with \(\tau_E\sim\lambda^{-1}\): a narrow Lorentzian spectrum retains environmental correlations for a longer time, whereas a broad spectrum rapidly loses its memory. 
The correlation-time argument and the BLP result therefore provide consistent characterizations of \(\lambda/\gamma\) as the reservoir-memory control parameter for the parameter family considered here.

\section{Collision-circuit construction, continuous-time limit, and validation}
\label{app:continuous_limit}

Here we give the explicit gate decomposition of the collision step in
Fig.~\ref{fig:collision_step} and derive its continuous-time limit.
We use the rotation convention:
\begin{align}
    R_{\alpha\alpha}^{ij}(\phi)
    &=
    \exp
    \left(
    -i\frac{\phi}{2}
    \sigma_\alpha^i\sigma_\alpha^j
    \right),
    &
    R_z^i(\phi)
    &=
    \exp
    \left(
    -i\frac{\phi}{2}Z_i
    \right).
    \label{eq:app_rotation_convention}
\end{align}
The three rotation angles appearing in the circuit are
\begin{align}
    \theta_{SP}
    &=
    g\Delta t
    =
    \sqrt{\frac{\gamma\lambda}{2}}\Delta t,
    &
    \theta_{PA}
    &=
    \sqrt{2\lambda\Delta t},
    &
    \theta_\delta
    &=
    \delta\Delta t.
    \label{eq:app_gate_angles}
\end{align}
Using
\(\sigma_+^S\sigma_-^P+\sigma_-^S\sigma_+^P
=(X_SX_P+Y_SY_P)/2\),
the system-pseudomode evolution is
\begin{align}
    U_{SP}(\Delta t)
    =
    R_{xx}^{SP}(\theta_{SP})
    R_{yy}^{SP}(\theta_{SP}).
    \label{eq:app_USP_gates}
\end{align}
The \(XX\) and \(YY\) generators commute for the same pair of qubits,
and hence this decomposition is exact.
The pseudomode decay is implemented by coupling \(P\) to an ancilla
through the collision Hamiltonian:
\begin{align}
    H_{PA}^{(c)}
    =
    \sqrt{\frac{2\lambda}{\Delta t}}
    \left(
    \sigma_+^P\sigma_-^A
    +
    \sigma_-^P\sigma_+^A
    \right),
    \label{eq:app_HPA_collision}
\end{align}
which gives
\begin{align}
    U_{PA}(\Delta t)
    =
    R_{xx}^{PA}(\theta_{PA})
    R_{yy}^{PA}(\theta_{PA}).
    \label{eq:app_UPA_gates}
\end{align}
The scaling
\(H_{PA}^{(c)}\propto\Delta t^{-1/2}\)
produces an excitation-loss probability of order \(\Delta t\) in
each collision.
The detuning evolution generated by
\(H_\delta=-\delta n_P\) is, up to a global phase,
\begin{align}
    U_\delta(\Delta t)
    =
    R_z^P(\theta_\delta).
    \label{eq:app_Udelta_gate}
\end{align}

The complete one-step channel on the enlarged system \(S+P\) is
\begin{align}
    \mathcal{E}_{\Delta t}[\rho_{SP}]
    =
    \operatorname{Tr}_A
    \Bigl[
    U_{PA}U_{SP}U_\delta
    \left(
    \rho_{SP}
    \otimes
    \lvert0\rangle_A\langle0\rvert
    \right)
    U_\delta^\dagger
    U_{SP}^\dagger
    U_{PA}^\dagger
    \Bigr].
    \label{eq:app_one_step_channel}
\end{align}
After \(N\) collisions, the state is
\begin{align} 
\rho_{SP}(N\Delta t) = \mathcal{E}_{\Delta t}^{N} [\rho_{SP}(0)]. 
\end{align}
To derive the continuous-time generator, we first expand the coherent part of the step as 
\begin{align} 
U_{SP}U_\delta \rho U_\delta^\dagger U_{SP}^\dagger = \rho - i\Delta t [H_{SP}+H_\delta,\rho] + \mathcal{O}(\Delta t^2). 
\label{eq:app_coherent_expansion} 
\end{align}

Since the ancilla is initialized in \(\lvert0\rangle_A\), the pseudomode-ancilla collision induces the Kraus operators:
\begin{align} 
    K_0
    &= {}_A\langle0|U_{PA}|0\rangle_A  
    \nonumber\\ 
    &= \lvert0\rangle_P\langle0\rvert + \cos\theta_{PA} \lvert1\rangle_P\langle1\rvert, 
    \nonumber\\
    K_1 
    &= {}_A\langle1|U_{PA}|0\rangle_A 
    \nonumber\\ 
    &= -i\sin\theta_{PA} \lvert0\rangle_P\langle1\rvert. \label{eq:app_PA_kraus} 
\end{align}
Using \(\theta_{PA}=\sqrt{2\lambda\Delta t}\) from Eq.~\eqref{eq:app_gate_angles}, the trigonometric factors in Eq.~\eqref{eq:app_PA_kraus} have the small-\(\Delta t\) expansions:
\begin{align}
    \sin^2\theta_{PA}
    &=
    2\lambda\Delta t+\mathcal{O}(\Delta t^2),
    \nonumber\\
    \cos\theta_{PA}
    &=
    1-\lambda\Delta t+\mathcal{O}(\Delta t^2).
    \label{eq:app_PA_trigonometric_expansion}
\end{align}
Using the Kraus operators in Eq.~\eqref{eq:app_PA_kraus} together with the expansions in Eq.~\eqref{eq:app_PA_trigonometric_expansion}, the two Kraus contributions become
\begin{align}
    K_0\rho K_0^\dagger
    &=
    \rho
    -
    \lambda\Delta t
    \left(
    n_P\rho+\rho n_P
    \right)
    +
    \mathcal{O}(\Delta t^2),
    \nonumber\\
    K_1\rho K_1^\dagger
    &=
    2\lambda\Delta t\,
    \sigma_-^P\rho\sigma_+^P
    +
    \mathcal{O}(\Delta t^2).
    \label{eq:app_PA_Kraus_contributions}
\end{align}
Summing the two contributions in Eq.~\eqref{eq:app_PA_Kraus_contributions} gives the pseudomode-ancilla collision map:
\begin{align}
    \mathcal{E}_{PA,\Delta t}[\rho] = \rho + 2\lambda\Delta t\, \mathcal{D}[\sigma_-^P]\rho + \mathcal{O}(\Delta t^2),
    \label{eq:app_PA_map_expansion}
\end{align}
and hence
\begin{align} 
    \frac{ \mathcal{E}_{PA,\Delta t}[\rho]-\rho }{\Delta t} = 2\lambda \mathcal{D}[\sigma_-^P]\rho + \mathcal{O}(\Delta t). 
    \label{eq:app_PA_limit} 
\end{align}
Combining the coherent expansion in Eq.~\eqref{eq:app_coherent_expansion} with the collision-map expansion in Eq.~\eqref{eq:app_PA_map_expansion} gives
\begin{align}
    \frac{
    \mathcal{E}_{\Delta t}[\rho]-\rho
    }{\Delta t}
    =
    -i[H_{SP}+H_\delta,\rho]
    +
    2\lambda
    \mathcal{D}[\sigma_-^P]\rho
    +
    \mathcal{O}(\Delta t).
    \label{eq:app_full_liouvillian}
\end{align}
Thus, taking \(\Delta t\rightarrow0\) in Eq.~\eqref{eq:app_full_liouvillian} reproduces the pseudomode master equation in Eq.~\eqref{eq:pseudomode_master}.

In addition, we validate the discrete collision circuit by comparing it with the
continuous pseudomode dynamics under the parameters used in the main text.
For an initially excited target system and a ground-state pseudomode, the single-excitation amplitudes satisfy \(\dot c_S=-igc_P\) and \(\dot c_P=-(\lambda-i\delta)c_P-igc_S\).
The analytical target-system amplitude is given in Eq.~\eqref{eq:BLP_cS}, and the corresponding excited-state population is \(P_e^S(t)=|c_S(t)|^2\).

\begin{figure*}
    \centering
    \includegraphics[width=0.9\textwidth]
    {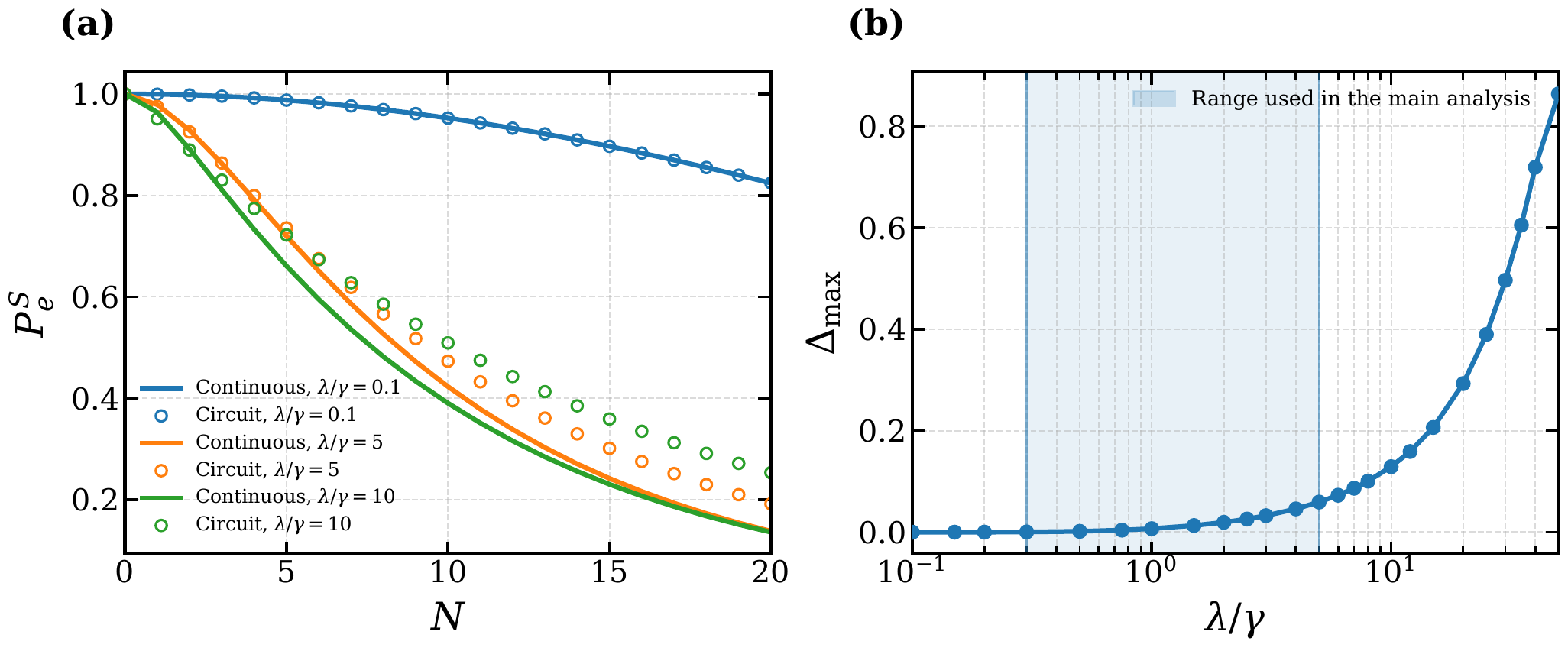}
    \caption{
    Validation of the discrete collision circuit against the
    continuous pseudomode model.
    (a) Excited-state populations for
    \(\lambda/\gamma=0.1\), \(5\), and \(10\).
    Solid curves show the continuous solution, and markers show the
    collision-circuit results.
    The target system and pseudomode are initially prepared in
    \(\lvert1\rangle_S\) and \(\lvert0\rangle_P\), respectively.
    (b) Maximum population deviation
    \(\Delta_{\max}(\lambda/\gamma)\) over
    \(N_{\max}=20\) collision steps across the extended spectral-width range.
    The shaded region denotes the range \(0.3\leq\lambda/\gamma\leq5\) used in the main analysis. The parameters are \(\gamma\Delta t=0.1\) and \(\delta=0\).
    }
    \label{fig:app_full_JC_validation}
\end{figure*}

Figure~\ref{fig:app_full_JC_validation}(a) compares the continuous
solution with the collision-circuit result for
\(\lambda/\gamma=0.1\), \(5\), and \(10\), representing increasingly broad reservoir spectra.
To quantify the discretization error over the full spectral-width
range, we evaluate
\begin{align}
    \Delta_{\max}\left(\frac{\lambda}{\gamma}\right)
    =
    \max_{0\leq N\leq N_{\max}}
    \left|
    P_{e,\mathrm{circ}}(N\Delta t;\lambda)
    -
    P_{e,\mathrm{cont}}(N\Delta t;\lambda)
    \right|.
    \label{eq:app_max_population_error}
\end{align}
As shown in Fig.~\ref{fig:app_full_JC_validation}(b), the deviation remains comparatively small over the main-analysis range \(0.3\leq\lambda/\gamma\leq5\) and increases toward broader reservoir spectra.
This increase occurs because the collision angle \(\theta_{PA}=\sqrt{2\lambda\Delta t}\) is no longer small at large \(\lambda/\gamma\).
We therefore take \(\lambda/\gamma=5\) as the upper boundary of the main analysis, within which the discrete circuit reliably approximates the continuous pseudomode dynamics.
The lower boundary \(\lambda/\gamma=0.3\) is instead set by the requirement that the prescribed cooling threshold be reached within \(N_{\max}=20\).

\section{Geometric length of the sequential collision circuit} 
\label{app:geometric_cost} 
Here we derive the geometric length of the specified sequential collision circuit shown in Figs.~\ref{fig:collision_step} and \ref{fig:geometric_channel_cost}. 
The construction follows the dilation-based implementation-dependent channel complexity of Ref.~\cite{Acevedo:2026:GeometricChannels}, which extends Nielsen's geometric formulation of unitary circuit complexity~\cite{Nielsen2006}. 
The geometric system is \(\widetilde{S}=S+P\), while the fresh ancilla \(A\) is treated as its environment. 
The total Hilbert space of \(S+P+A\) has dimension \(8\), so the traceless generator space is \(\mathfrak{su}(8)\), with dimension \(8^2-1=63\). 

Let \(\{P_\mu\}_{\mu=1}^{63}\) denote the traceless three-qubit Pauli strings, which satisfy \(\operatorname{Tr}(P_\mu^\dagger P_\nu)=8\delta_{\mu\nu}\).
We introduce the Hilbert--Schmidt-orthonormal basis \(E_\mu=P_\mu/\sqrt{8}\).
A traceless Hermitian generator \(H(s)\) can then be expanded as \begin{align}
    H(s) = \sum_{\mu=1}^{63} h_\mu(s)E_\mu, 
    \label{eq:app_generator_expansion}
\end{align} 
where \(h_\mu(s)\in\mathbb{R}\) is the expansion coefficient associated with the normalized Pauli-string direction \(E_\mu\).

The penalty matrix \(\Omega\) assigns different geometric costs to different operator directions, encoding which controls or couplings are regarded as more expensive. 
In a basis that diagonalizes \(\Omega\), its diagonal entries give the penalty weights associated with the corresponding generator directions~\cite{Acevedo:2026:GeometricChannels}. 
The normalized geometric norm is defined by 
\begin{align}
    \|H(s)\|_{\Omega} = \sqrt{ \frac{ \mathbf{h}(s)^\dagger \Omega \mathbf{h}(s) }{ \operatorname{Tr}\Omega }}, 
    \label{eq:app_weighted_norm}
\end{align}
where \(\mathbf{h}(s)=(h_1(s),\ldots,h_{63}(s))^{\mathsf T}\). 

For a piecewise-smooth unitary path \(U(s)\), the right-trivialized Hermitian generator is
\begin{align} 
    H(s) = i\dot U(s)U^\dagger(s). 
\label{eq:app_right_generator} 
\end{align} 
The dilation-based channel-path length is
\begin{align} 
    L_{\Omega}[U] = \int ds\, \left\| H(s)-\mathcal{P}_A[H(s)] \right\|_{\Omega}.
    \label{eq:app_channel_path_length} 
\end{align}
Here,
\begin{align}
    \mathcal{P}_A[X] = \frac{I_{SP}}{4} \otimes \operatorname{Tr}_{SP}[X] 
    \label{eq:app_ancilla_projection}
\end{align}
extracts the component of \(X\) that acts only on the ancilla. 
Such an ancilla-local component does not affect the reduced channel on \(S+P\) and is therefore excluded from the channel-path length ~\cite{Acevedo:2026:GeometricChannels}.

As shown in Fig.~\ref{fig:collision_step}, the specified sequential circuit consists of the detuning, system-pseudomode, and pseudomode-ancilla unitary segments. 
The corresponding elementary generators are proportional to \(Z_P\), \(X_SX_P\), \(Y_SY_P\), \(X_PX_A\), and \(Y_PY_A\), where identity operators on qubits not shown are implicit.
The generators \(Z_P\), \(X_SX_P\), and \(Y_SY_P\) act within the geometric system \(S+P\), whereas \(X_PX_A\) and \(Y_PY_A\) describe system-environment interactions between the pseudomode and ancilla. 
The latter are therefore retained in the implementation-dependent cost.
Every generator contains a traceless Pauli operator acting on at least one qubit in \(S+P\), so its partial trace over \(S+P\) vanishes. 
Explicitly,
\begin{align} 
    \operatorname{Tr}_{SP} \left[ I_S\otimes Z_P\otimes I_A \right] 
    &= \operatorname{Tr}(I_S) \operatorname{Tr}(Z_P)I_A = 0, \nonumber\\
    \operatorname{Tr}_{SP} \left[ X_S\otimes X_P\otimes I_A \right] 
    &= \operatorname{Tr}(X_S) \operatorname{Tr}(X_P)I_A = 0, \nonumber\\
    \operatorname{Tr}_{SP} \left[ Y_S\otimes Y_P\otimes I_A \right] 
    &= \operatorname{Tr}(Y_S) \operatorname{Tr}(Y_P)I_A = 0, \nonumber\\
    \operatorname{Tr}_{SP} \left[ I_S\otimes X_P\otimes X_A \right] 
    &= \operatorname{Tr}(I_S) \operatorname{Tr}(X_P)X_A = 0, \nonumber\\ 
    \operatorname{Tr}_{SP} \left[ I_S\otimes Y_P\otimes Y_A \right] 
    &= \operatorname{Tr}(I_S) \operatorname{Tr}(Y_P)Y_A = 0. 
\label{eq:app_zero_partial_traces} 
\end{align} 
It follows from Eqs.~\eqref{eq:app_ancilla_projection} and \eqref{eq:app_zero_partial_traces} that 
\begin{align} 
    \mathcal{P}_A[H_j] = 0 
    \label{eq:app_zero_environment_projection}
\end{align} 
for every segment \(j\) of the specified circuit path. 
Therefore, Eq.~\eqref{eq:app_channel_path_length} reduces to the ordinary geometric length of each implemented segment.
Ancilla reset and fresh-ancilla preparation between successive collisions are not included in the unitary path length evaluated here, because they are not unitary segments of the specified collision path \(U_{PA}U_{SP}U_\delta\).

As an analytically convenient baseline, we now specialize to the equal-penalty Hilbert--Schmidt geometry by setting \(\Omega=I_{63}\).
This choice assigns the same cost to all 63 normalized Pauli-string generator directions. 
Because the basis \(\{E_\mu\}\) is Hilbert--Schmidt orthonormal, Eq.~\eqref{eq:app_weighted_norm} reduces to 
\begin{align} 
    \|H(s)\|_{\Omega} = \frac{1}{\sqrt{63}} \|H(s)\|_{\rm HS}.
    \label{eq:app_HS_normalized_norm} 
\end{align} 
For an unnormalized three-qubit Pauli string \(P_\mu\), 
\begin{align}
    \operatorname{Tr} \left( P_\mu^\dagger P_\mu \right) = 8, \qquad \|P_\mu\|_{\rm HS} = \sqrt{8}.
    \label{eq:app_pauli_norm}
\end{align}
Consider an elementary rotation: 
\begin{align}
U_\mu(s) = \exp \left( -is\frac{\theta}{2}P_\mu \right), \qquad 0\leq s\leq1. 
\label{eq:app_elementary_path} 
\end{align}
Using Eq.~\eqref{eq:app_right_generator}, the right-trivialized generator of the path in Eq.~\eqref{eq:app_elementary_path} is 
\begin{align} 
    H_\mu = \frac{\theta}{2}P_\mu. 
    \label{eq:app_elementary_generator}
\end{align}
Using Eqs.~\eqref{eq:app_HS_normalized_norm}, \eqref{eq:app_elementary_generator}, and \eqref{eq:app_pauli_norm}, its geometric length is
\begin{align} 
    L(\theta) 
    &= \int_0^1 ds\, \frac{1}{\sqrt{63}} \left\| \frac{\theta}{2}P_\mu \right\|_{\rm HS} 
    \nonumber\\
    &= \frac{1}{\sqrt{63}} \frac{|\theta|}{2} \sqrt{8}
    \nonumber\\
    &= \sqrt{\frac{2}{63}} |\theta|. 
    \label{eq:app_single_rotation_length}
\end{align}

The sequential circuit contains one detuning rotation with angle \(\theta_\delta\), two system-pseudomode rotations with angle \(\theta_{SP}\), and two pseudomode-ancilla rotations with angle \(\theta_{PA}\). 
The length of a piecewise path is the sum of the integrals over its successive segments.
Applying Eq.~\eqref{eq:app_single_rotation_length} to the five elementary rotations gives
\begin{align} 
L_\delta &= \sqrt{\frac{2}{63}} |\theta_\delta|, \nonumber\\ L_{SP} &= 2\sqrt{\frac{2}{63}} |\theta_{SP}| = \sqrt{\frac{8}{63}} |\theta_{SP}|, \nonumber\\ L_{PA} &= 2\sqrt{\frac{2}{63}} |\theta_{PA}| = \sqrt{\frac{8}{63}} |\theta_{PA}|. 
\label{eq:app_segment_lengths} 
\end{align}
Summing the segment lengths in Eq.~\eqref{eq:app_segment_lengths}, the one-step geometric length is
\begin{align} 
\ell_{\rm PM}^{\rm seq} &= L_\delta+L_{SP}+L_{PA} \nonumber\\ &= \sqrt{\frac{8}{63}} \left( |\theta_{SP}|+|\theta_{PA}| \right) + \sqrt{\frac{2}{63}} |\theta_\delta|. 
\label{eq:app_ellPM_angles}
\end{align}
Substituting the circuit angles in Eq.~\eqref{eq:app_gate_angles} into Eq.~\eqref{eq:app_ellPM_angles} gives
\begin{align} 
    \ell_{\rm PM}^{\rm seq}(\lambda) = \sqrt{\frac{8}{63}} \left[ \sqrt{\frac{\gamma\lambda}{2}}\Delta t + \sqrt{2\lambda\Delta t} \right] + \sqrt{\frac{2}{63}} |\delta|\Delta t. 
\end{align}
This reproduces Eq.~\eqref{eq:ellPM_lambda}.
The additivity used above follows from splitting the path-length integral over the consecutive circuit segments; it does not require generators belonging to different segments to commute.

\section{Minimum excitation within the observation window}
\label{app:min_population}

The cooling-step count \(N_{\rm cool}\), defined in Eq.~\eqref{eq:Ncool_and_Pe}, characterizes the number of collision steps required for the target to reach the prescribed cooling threshold.
However, it is undefined when the threshold is not reached within the finite observation window, as occurs for several parameter values in the hardware implementation. 
To provide a complementary measure of finite-step cooling performance, we therefore evaluate the minimum excited-state population:
\begin{align}
    P_e^{\min}(\lambda)
    =
    \min_{0\leq N\leq N_{\max}}
    P_e(N\Delta t;\lambda).
    \label{eq:Pe_min}
\end{align}
This quantity allows all spectral widths to be compared over the same fixed number of collision steps, independently of whether the cooling threshold is reached.

\begin{figure*}
    \centering
    \begin{minipage}{0.45\textwidth}
        \centering
        \includegraphics[width=\linewidth]
        {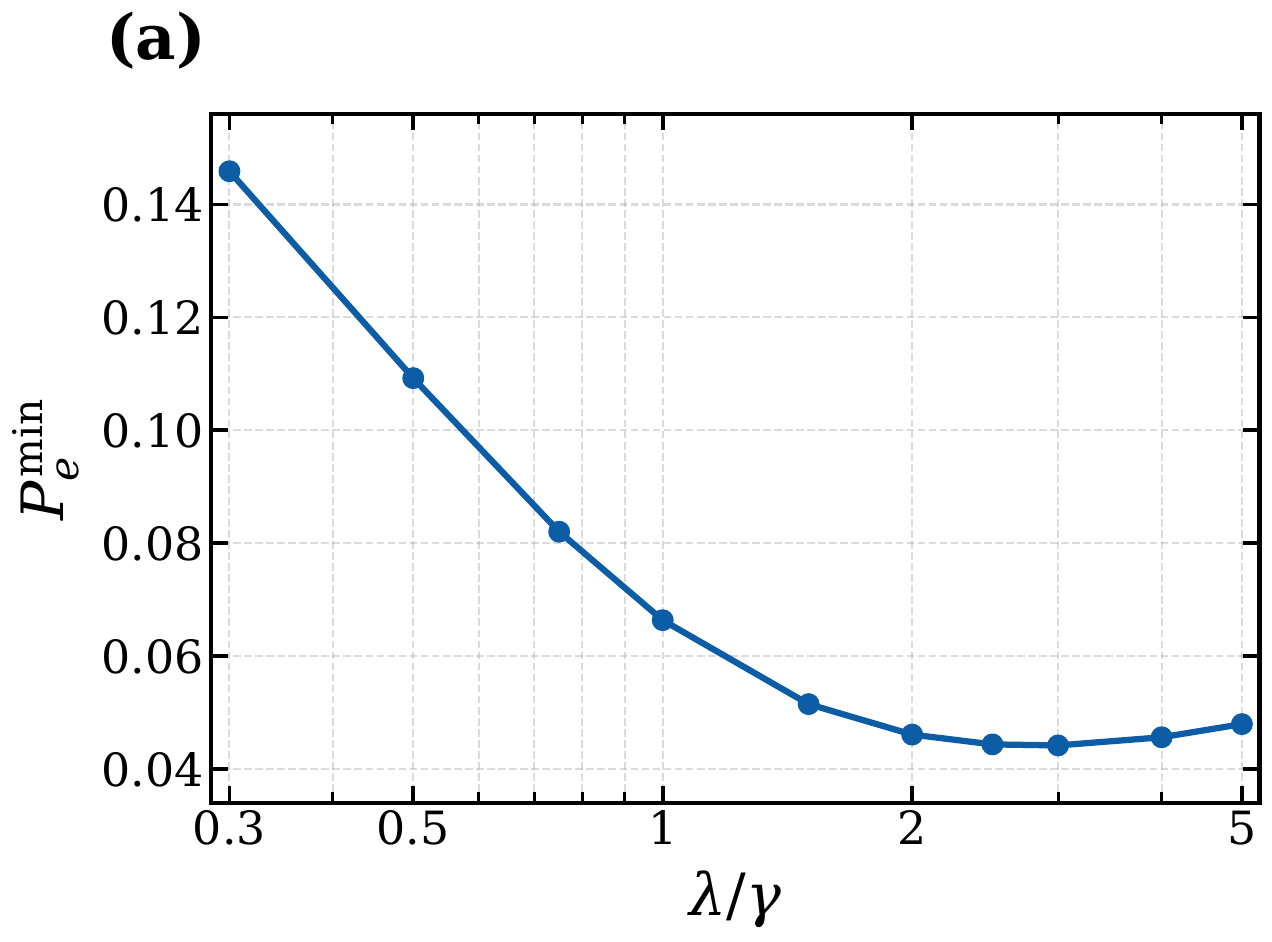}
    \end{minipage}
    \begin{minipage}{0.45\textwidth}
        \centering
        \includegraphics[width=\linewidth]
        {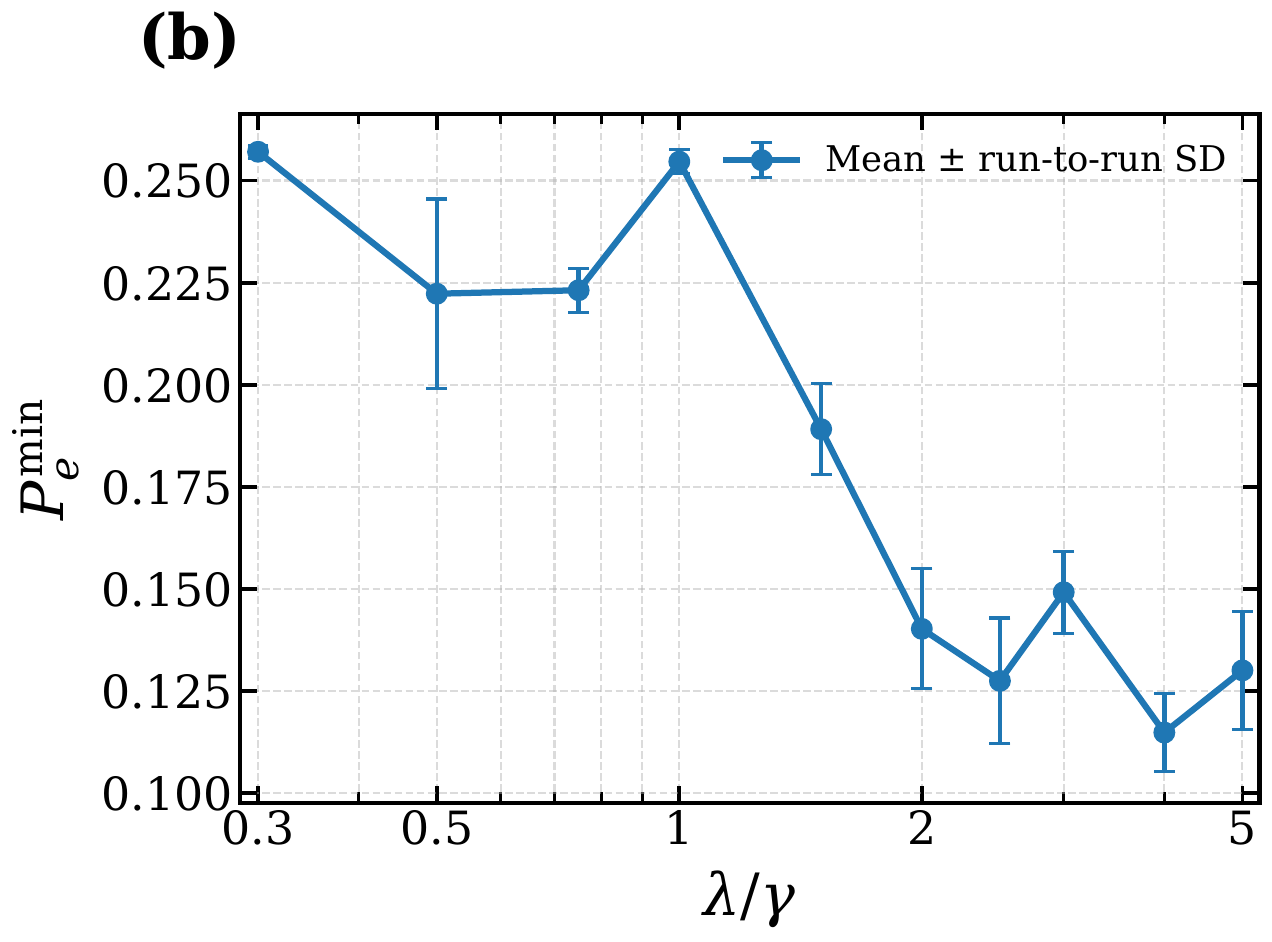}
    \end{minipage}
    \caption{
    Minimum excited-state population attained within \(N_{\max}=20\) collision steps as a complementary measure of finite-step cooling performance:
    (a) matrix-based simulation and
    (b) hardware implementation.
    Both panels use the same spectral-width range, \(0.3\leq\lambda/\gamma\leq5\), as the main analysis.
    The hardware points show the mean over three independent runs, and the error bars indicate the run-to-run standard deviation.
    Although neither result is strictly monotonic over the entire range, lower values of \(P_e^{\min}\) are generally attained on the larger-\(\lambda/\gamma\), more Markovian side. The lines are guides for the eye.
    }
    \label{fig:min_Pe_comparison}
\end{figure*}

Figure~\ref{fig:min_Pe_comparison} shows
\(P_e^{\min}(\lambda)\) for the matrix-based simulation and the hardware implementation.
For the hardware results, the points and error bars denote the mean and run-to-run standard deviation over three independent runs, respectively.
The matrix-based result is not strictly monotonic over the entire spectral-width range. 
Nevertheless, \(P_e^{\min}\) is generally lower on the larger-\(\lambda/\gamma\), more Markovian side than on the narrow-band, more non-Markovian side. 
Although hardware noise and run-to-run fluctuations prevent a monotonic dependence from being resolved, the hardware result shows the same broad contrast, with lower values of \(P_e^{\min}\) generally attained on the larger-\(\lambda/\gamma\) side. 
Thus, for the finite-step protocol and parameter range considered here, the improved cooling performance on the more Markovian side is not solely a consequence of defining performance through the threshold-crossing step \(N_{\rm cool}\).

\bibliography{reference}

\end{document}